\documentclass[twocolumn,nofootinbib,amsmath,amssymb,aps,prd,balancelastpage,superscriptaddress]{revtex4-1}
\usepackage{color}
\usepackage{xcolor}
\usepackage[active]{srcltx}
\usepackage{amsmath,amsfonts,amssymb,amsthm,amstext,amscd,eucal,srcltx}
\usepackage{epsfig,graphicx,bm}
\usepackage{epstopdf, epsf}
\usepackage{dcolumn}
\usepackage{comment}
\usepackage{hyperref}
\usepackage{float}
\usepackage{lipsum}
\usepackage{lineno}
\usepackage[caption=false]{subfig}

\newcommand{\be}{\begin{equation}}
\newcommand{\ee}{\end{equation}}

\newcommand{\bse}{\begin{subequations}}
\newcommand{\ese}{\end{subequations}}
\newcommand{\bea}{\begin{eqnarray}}
\newcommand{\eea}{\end{eqnarray}}
\newcommand{\ba}{\begin{array}}
\newcommand{\ea}{\end{array}}
\newcommand{\bc}{\begin{center}}
\newcommand{\ec}{\end{center}}

\begin{document}

%\linenumbers

\preprint{IPM/P-2012/009}  %\cr
          %\eprint{arXiv:1202.nnnn}}
\vspace*{3mm}

\title{First Simultaneous  K$^-$p $\rightarrow (\Sigma^0/\Lambda) \, \pi^0$ Cross Sections Measurements at 98~MeV/c}%

%VIP-2 Lead experiment sets the strongest bounds on Non-Commutative Quantum Gravity Models in atomic physics

%Strongest bounds in atomic physics on Non-Commutative Quantum Gravity from VIP-2 Lead

\author{{Kristian Piscicchia$^{a,b}$, Magdalena Skurzok$^{*,c,d}$, Michael Cargnelli$^{e,b}$, Raffaele Del Grande$^{f,b}$, Laura Fabbietti$^{g,f}$, Johann Marton$^{e,b}$, Pawel Moskal$^{c,d}$, Alessandro Scordo$^b$, Àngels Ramos$^h$, Diana Laura Sirghi$^{b,i}$, Oton Vazquez Doce$^b$, Johann Zmeskal$^{e,b}$,
S\l{}awomir Wycech$^{j}$, 
Paolo Branchini$^{k}$,
Filippo Ceradini$^{l,k}$,
Eryk Czerwi\'nski$^{c,d}$,
Erika De Lucia$^b$,
Salvatore Fiore$^{m,n}$,
Andrzej Kupsc$^{o,j}$,
Giuseppe Mandaglio$^{p,q}$,
Matteo Martini$^{b,r}$,
Antonio Passeri$^{k}$,
Vincenzo Patera$^{s,n}$,
Elena Perez Del Rio$^{c,d}$,
Andrea Selce$^{m,b}$,
Micha\l{} Silarski$^{c,d}$
and Catalina Curceanu$^{b}$}\\ 
\vspace{0.5 cm}
{\it$^a$ Centro Ricerche Enrico Fermi - Museo Storico della Fisica e Centro Studi e Ricerche “Enrico Fermi”, Roma, Italy, EU}\\
{\it$^b$ Laboratori Nazionali di Frascati INFN, Frascati (Rome), Italy, EU}\\
{\it$^c$ Institute of Physics, Jagiellonian University, Kracow, Poland, EU}\\
{\it$^d$Center for Theranostics, Jagiellonian University, Krakow, Poland, EU}\\
%{magdalena.skurzok@uj.edu.pl}\\
{\it$^e$ Stefan-Meyer-Institute for subatomic physics, Austrian Academy of Science, Austria, EU}\\
{\it$^f$ Physik Department E62, Technische Universität München, Garching, Germany, EU}\\
{\it$^g$ Excellence Cluster ``Origin and Structure of the Universe'', Garching, Germany, EU\\}
{\it$^h$ Facultat de F\'isica, Universitat de Barcelona, Barcelona, Spain, EU}\\
{\it$^i$ IFIN-HH, Institutul National pentru Fizica si Inginerie Nucleara Horia Hulubei, Romania, EU}\\
{\it$^j$ National Centre for Nuclear Research, Warsaw, Poland, EU}\\
{\it$^k$ INFN Sezione di Roma Tre, Roma, Italy}\\
{\it$^l$ Dipartimento di Matematica e Fisica dell'Universit\`a ``Roma Tre'', Roma, Italy}\\
{\it$^m$ ENEA, Department of Fusion and Technology for Nuclear Safety and Security, Frascati (Rome), Italy}\\
{\it$^n$ INFN Sezione di Roma, Roma, Italy}\\
{\it$^o$Department of Physics and Astronomy, Uppsala University, Uppsala, Sweden}\\
{\it$^p$ Dipartimento di Scienze Matematiche e Informatiche, Scienze Fisiche e Scienze
della Terra dell'Universit\`a di Messina, Messina, Italy}\\
{\it$^q$ INFN Sezione di Catania, Catania, Italy}\\
{\it$^r$ Dipartimento di Scienze e Tecnologie applicate, Universit\`a ``Guglielmo Marconi'', Roma, Italy}\\
{\it$^s$ Dipartimento di Scienze di Base ed Applicate per l'Ingegneria dell'Universit\`a
``Sapienza'', Roma, Italy}\\
}
\thanks{magdalena.skurzok@uj.edu.pl}

\vspace{0.5 cm}

\begin{abstract}
\noindent
We report the first simultaneous and independent measurements of the  K$^{-}$p $\rightarrow \Sigma^0 \, \pi^{0}$ and K$^{-}$p $\rightarrow \Lambda \, \pi^{0}$  cross sections around 100~MeV/c kaon momentum. The kaon beam delivered by the DA$\Phi$NE collider was exploited to detect K$^-$ absorptions on Hydrogen atoms, populating the gas mixture of the KLOE drift chamber. The precision of the measurements
($\sigma_{K^- p \rightarrow \Sigma^0 \pi^0} = 42.8 \pm 1.5 (stat.) ^{+2.4}_{-2.0}(syst.) \ \mathrm{mb}$
and $\sigma_{K^- p \rightarrow \Lambda \pi^0} = 31.0 \pm 0.5 (stat.) ^{+1.2}_{-1.2}(syst.) \ \mathrm{mb}\,$) is the highest yet obtained in the low kaon momentum regime.
\end{abstract}

\maketitle

%\section{Introduction}

The experimental investigation of the $\overline{\mathrm{K}}$N low energy interaction aims at providing constrains to the non-perturbative QCD models in the strangeness sector. Chiral perturbation theory cannot be extended from SU(2) to SU(3) when a strangeness $S = -1$ quark is introduced in the theory. This is due to the existence of broad resonances ($\Lambda(1405)$ and $\Sigma(1385)$) just below the $\overline{\mathrm{K}}$N threshold. The $\Lambda(1405)$ is an isospin $I=0$ state whose existence was first predicted by Dalitz and Tuan \cite{DT1,DT2} as a resonant structure in a coupled-channel formulation of the $\overline{\mathrm{K}}$N interaction. Since its first experimental observation \cite{expl1405} the interpretation of this quantum state is still controversial, although the inability of the quark models to predict its mass correctly has strengthened the original idea of the $\Lambda(1405)$ being a meson-baryon quasi-bound state \cite{DT1,DT2}, an interpretation that was later formulated in terms of modern chiral lagrangians \cite{kaiser,oset-ramos}. The spectral shape of the resonance is observed to depend on both the adopted production mechanism and the measured decay channel \cite{exper,hemingway,zychor,moriya,agakishiev}, a fact that is naturally explained in terms of a two-poles structure for the $\Lambda(1405)$, first pointed out in \cite{oller} but found in all chiral unitary models, of which the most recent versions \cite{IHW,Guo,feijoo,lu} (see also the review papers of \cite{Hyodo} and \cite{Mai:2020ltx}) have already been constrained by the precise measurement of the energy shift and width of the 1s kaonic-hydrogen state from the SIDDHARTA collaboration \cite{SIDD}.
While all recent chiral models provide a rather similar description of the K$^{-}$p scattering length, the subthreshold elastic and inelastic amplitudes, which are necessary to describe kaon nuclear absorption processes, may differ by factors of more than three.  Since the extrapolation of the scattering amplitude below the threshold is so challenging, accurate experimental input on the K$^-$p$\rightarrow \Sigma^{0} \pi^{0}$ dynamics above the threshold is strongly needed. The $\Sigma^{0} \pi^{0}$ channel, being almost purely isospin $I=0$ (with a small $I=2$ component), gives much cleaner information with respect to the charged $\Sigma^{\pm}\pi^\mp$ channels \cite{piscicchiaepjweb2018}.

Presently, the available data for the inelastic K$^{-}$p $\rightarrow \Sigma^{0} \pi^{0}$ cross section close to threshold are obtained by means of indirect extrapolations, in bubble chamber experiments (\cite{s0pi01,martin,ferrari} and references therein), from the measurement of the $\Lambda \pi^0$ channel, by exploiting the isospin symmetry. Three cross sections are given at mean kaon momenta   $p_\mathrm{{K^{-}}} = 120, 160$ and 200~MeV/c, which are affected by sizable relative errors (as large as 50\% at $p_\mathrm{{K^{-}}} = 120$~MeV/c), and where the $p_\mathrm{{K^{-}}}$ momentum intervals around the central values are 25~MeV/c wide.   
%and an uncertainty on   $p_\mathrm{{K^{-}}}$ of 12.5~MeV/c.

This work takes advantage of the low momentum negatively charged kaon beam produced at the DA$\Phi$NE e$^+$e$^-$ collider ~\cite{Gallo}, which has been designed to work at the center of mass energy of the $\phi$ meson, thereby delivering almost monochromatic charged kaons with $p_\mathrm{{K^{-}}}\sim$127~MeV/c. Utilizing the KLOE~\cite{KLOEcite} spectrometer as an active detector  $\Sigma^{0} \pi^{0}$ and $\Lambda \pi^{0}$ events are reconstructed. The K$^-$ absorption in-flight on Hydrogen atoms, from the gas mixture of the KLOE Drift Chamber, is disentangled to independently obtain the K$^{-}$p $\rightarrow (\Sigma^{0}/\Lambda) \, \pi^{0}$ cross sections at $p_\mathrm{{K^{-}}}= 98 \pm 10$~MeV/c. \newline

%(a mixture of 90\% in volume of Helium and 10\% in volume of Isobutane C$_4$H$_{10}$)
%The paper is organized as follow. In Section 2 the $\Sigma^{0} \pi^{0}$ events selection is described. Section 3 is dedicated to present the kinematic distributions of the simulated processes which contribute to the measured shapes. In Section 4 selection criteria of the in-flight K$^{-}$H $\rightarrow \Sigma^{0} \pi^{0}$ events are presented. The fit of the measured distributions to the simulated processes is shown in Section 5. The results of the analysis and comparisons with the previous experimental and theoretical findings are outlined in Sections 6 and 7. In Section 8 the conclusions are given.

%\section{Data Sample}

In this work a total integrated luminosity of 1.74~fb$^{-1}$, corresponding to the 2004/2005 KLOE data taking campaign, was analyzed by the AMADEUS collaboration.
The KLOE detector is centered around the DA$\Phi$NE interaction region and consists of a large cylindrical Drift Chamber (DC)~\cite{KLOEdc}, for tracking analysis, and a fine sampling lead-scintillating fibers calorimeter~\cite{KLOEemc}, used in this study
to determine position, time and energy of the
ionization deposits in the fibers
caused by particles hitting the calorimeter (clusters), all immersed in the axial magnetic field of a superconducting solenoid (0.52 T). The DC has an inner radius of 0.25~m, an outer radius of 2~m and a length of 3.3~m and is filled with a mixture of Helium and Isobutane C$_4$H$_{10}$ (90\% of Helium and 10\% of Isobutane in volume).
For more details on the DC and calorimeter performances, resolutions for the tracks and clusters reconstruction, we refer to Refs. \cite{KLOEdc,KLOEemc}. 
The strategy for the measurement of the K$^-$p $\rightarrow \Sigma^0 \pi^0$ and K$^-$p $\rightarrow \Lambda \pi^0$ cross sections consists in the identification of K$^-$ absorptions in-flight on the Hydrogen atoms of the Isobutane molecules.  \newline

%The analysed data include both K− captures at-rest and in-flight. In the first case the negatively charged kaon is slowed down in the materials of the detector and consequently absorbed in a highly excited atomic orbit, from which it cascades down till it is absorbed by the nucleus through the strong interaction. In the latter case the kaon penetrates the electronic cloud and interacts with the nucleus with an average momentum of about 100 MeV/c.

%\section{events selection}\label{sec3}

The identification of a $\Lambda(1116)$ hyperon, through its decay into a proton and a negatively charged pion (BR$_{\Lambda \rightarrow p \pi^{-}}$ = (63.9~$\pm$~0.5)~\%) \cite{PDG}, represents the signature of a hadronic interaction and the starting point of this analysis.
Protons and pions are selected by combining $dE/dx$ (truncated mean of the ADC collected counts due to ionization in the DC gas) versus momentum information provided by the DC, with the measurement of the clusters energies, as described in Ref.~\cite{acta,oton2016}. K$^-$ nuclear absorptions in the gas filling the DC volume are sorted by requiring the radial distance of the $\Lambda$ decay vertex from the DA$\Phi$NE beam pipe axis ($\rho_{\Lambda}$) to be greater than $30$~cm. The cut is optimized based on Monte Carlo (MC) simulations (performed by means of the standard KLOE GEANT digitization (GEANFI \cite{recons})), in order to minimize the contamination with K$^-$ absorptions in the DC entrance wall (aluminated carbon fiber), which amounts to less then 3\% for the selected sample.

The events selection proceeds through the photons identification by time of flight. Following the convention $\pi^0 \rightarrow \gamma_1 \, \gamma_2$ and $\Sigma^0 \rightarrow \Lambda \gamma_3$, for the two channels under study,  the two (or three) neutral clusters in the calorimeter (characterized by energy $E>20$~MeV, no associated track and not produced in a K$^+$ decay) are first selected by minimizing a time of flight based $\chi_t^2$ variable. Further, the photons are associated to the $\pi^0$ (and to the $\Sigma^0$) by a second $\chi_m^2$ analysis, using the $m_{\gamma_1\gamma_2}$ or the combined $m_{\gamma_1\gamma_2}$ and $m_{\Lambda\gamma_3}$ invariant masses information respectively.
The constraints $\chi_t^2 < 20$, $\chi^2_{m_{\gamma_1\gamma_2}} < 5$ and $\chi^2_{m_{\Lambda\gamma_3}} < 4$ are tuned based on MC studies. 
A check against cluster splitting (single clusters in the calorimeter erroneously
recognized as two clusters) is performed by analyzing the clusters distance as a function of their energy~\cite{acta}. The selected sample is not significantly affected by splitting.
The algorithm's efficiency for the photons clusters recognition is $0.98\pm0.01$. 

\begin{figure}[h!]
\begin{center}
\includegraphics[width=8.0cm,height=6.0cm]{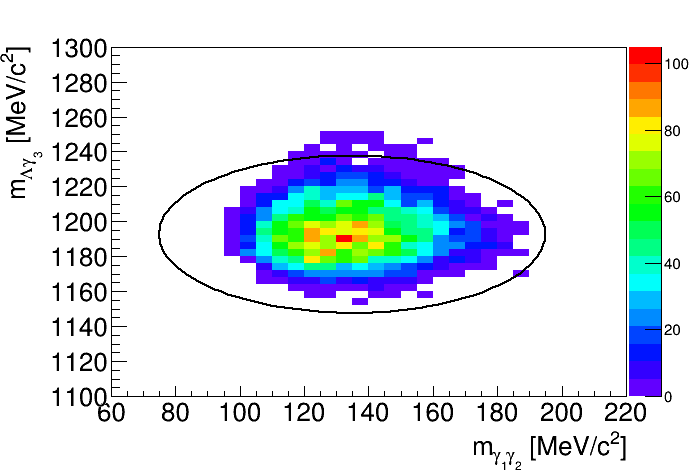}
\caption{The plot shows the $m_{\Lambda \gamma_{3}}$ distribution as function of $m_{\gamma_{1}\gamma_{2}}$. The black ellipse represents the applied selection in the two invariant masses space.}
\label{elypsis}
\end{center}
\end{figure}

In Fig.~\ref{elypsis} $m_{\Lambda \gamma 3}$ is represented as function of $m_{\gamma 1\gamma 2}$. Gaussian fits to the invariant mass distributions yield the resolutions $\sigma_{\gamma_1 \gamma_2}\sim$ 20~MeV/c and $\sigma_{\Lambda \gamma_3}\sim$  15~MeV/c respectively. Accordingly, the following additional cuts are applied for the two channels: $\left(m_{\Lambda \gamma_3} - m_{\Sigma^0} \right)^2/(3\sigma_{\Lambda \gamma_3})^{2} + \left(m_{\gamma_1 \gamma_2} - m_{\pi^0} \right)^2/(3\sigma_{\gamma_1 \gamma_2})^{2} \, < \, 1$ and  $\left| m_{\gamma_1 \gamma_2} - m_{\pi^0} \right| \, < \, 3\sigma_{\gamma_1 \gamma_2}$. \newline

%The optimization of the following phase space selections, aimed to pin down K$^-$H $\rightarrow (\Sigma^{0}/\Lambda)\, \pi^{0}$ captures in-flight, taking advantage of the distinctive kinematics of these processes, requires simulations of the involved reactions. The calculated distributions will then also serve for the data fit.

The optimization of the following phase space selections is aimed to disentangle the signal processes (K$^-$H $\rightarrow (\Sigma^{0}/\Lambda)\, \pi^{0}$ absorptions in-flight) from the competing background reactions, taking advantage of the signal's distinctive kinematics. This requires simulations of the involved reactions, which also serve for the data fit.

When a negatively charged kaon is absorbed in the DC gas, the final state $\Sigma^{0} \pi^{0}$ can be produced in K$^-$ + H $\rightarrow \Sigma^{0} + \pi^{0}$ at-rest ($ar$) or in-flight ($if$), or because the K$^-$ is absorbed on a bound proton in $^{4}\hspace{-0.03cm}$He or $^{12}\hspace{-0.03cm}$C, $ar$ or $if$. 
Analogous reactions also give rise to the $\Lambda \pi^{0}$ production both in direct processes (K$^-$p $\rightarrow \Lambda \pi^{0}$) or as a result of a $\Sigma^0$ decay (K$^-$p $\rightarrow \Sigma^0 \pi^{0}\rightarrow \Lambda \gamma \pi^0$).
Besides the direct processes and $\Sigma^0$ decay, in the nucleus also elastic or inelastic Final State Interactions (FSI) can end up with the observed hyperon ($Y \, = \, \Sigma^0 \, ,\, \Lambda$) and $\pi^{0}$. After phase space selections the contribution of the FSI is minor, and will be considered in the evaluation of the systematic uncertainties. 

The case of K$^-$ absorption on H is straightforward, since the kinematics is determined by energy-momentum conservation, for both $ar$ and $if$. Calculation of the $if$ reaction requires as input the negatively charged kaon momentum, which is sampled according to the true MC (i.e. not passed for the events reconstruction) momentum distribution of the kaons inside the DC volume.
The simulations of the K$^-$ nuclear absorption processes are performed according to a phenomelogical model~\cite{io,PisWycCur,lpi,lproton} which provides, for each reaction, the total hyperon-pion momentum probability distribution P($p_{Y\pi^{0}}$). The kinematics of an event is completely defined by the momentum vectors ($\textbf{\textit{p}}_{Y}, \textbf{\textit{p}}_{\pi^{0}}$) (the residual(s) nucleus track(s) is not detected), which are generated by sampling the P($p_{Y\pi^{0}}$) distribution, and applying energy and momentum conservation. For each event the calculated ($\textbf{\textit{p}}_{Y}, \textbf{\textit{p}}_{\pi^{0}}$) pairs represent the input for the KLOE GEANT digitization, followed by the event reconstruction.

\begin{figure} [h!]
\centering
\includegraphics[width=8.0cm, height=6.0cm]{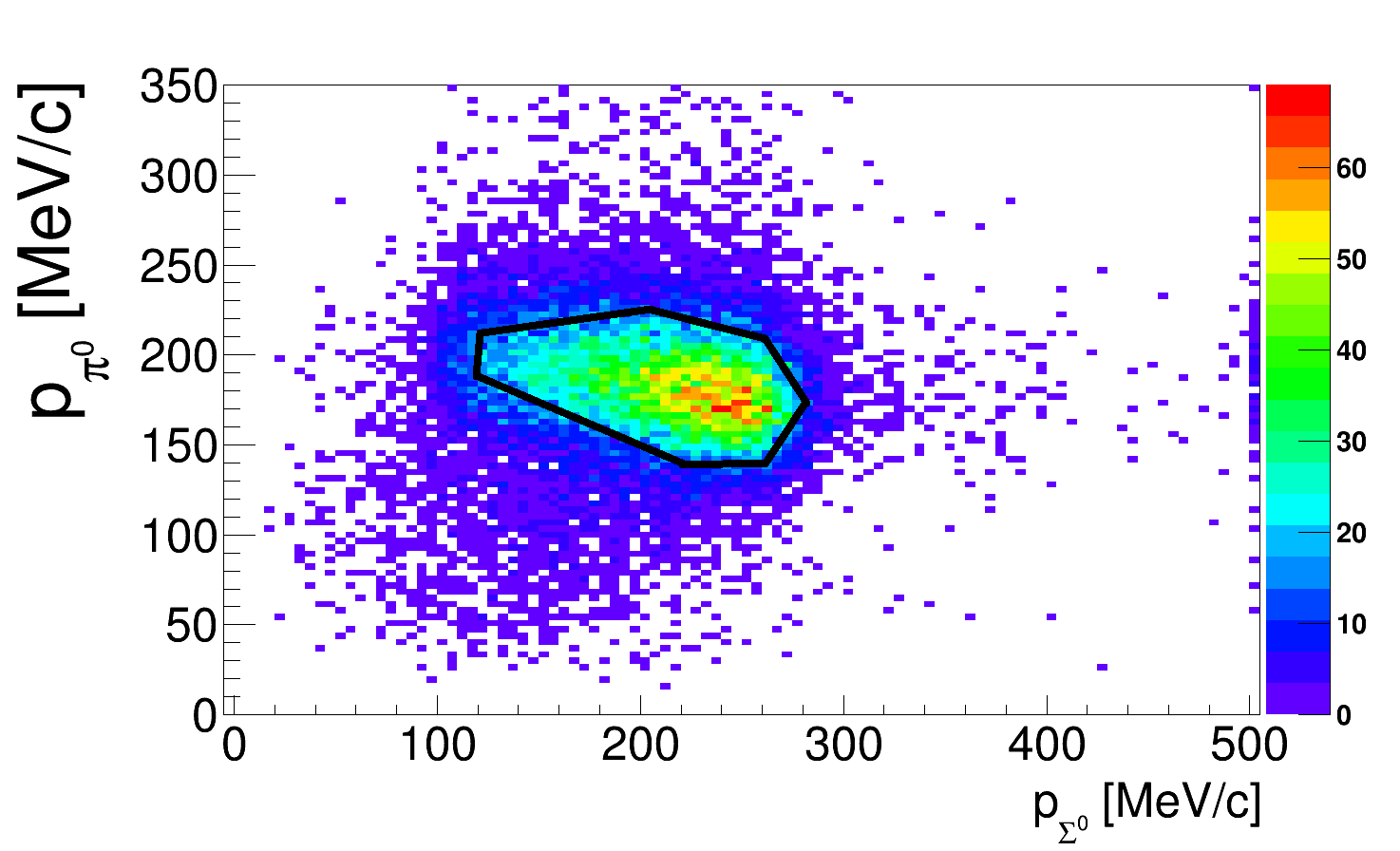}\\
\includegraphics[width=8.0cm, height=6.0cm]{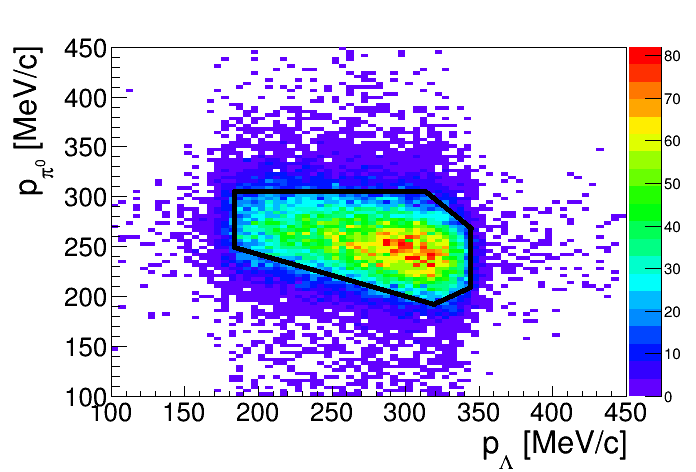}
%\vspace{-1.0cm}
\caption{The plot shows reconstructed MC $p_{\pi^{0}}$ vs. $p_{Y}$ distributions for the K$^-$ \, H $\rightarrow \Sigma^{0} \, \pi^{0}$ $if$ reaction (top) and K$^-$ \, H $\rightarrow \Lambda \, \pi^{0}$ $if$ reaction (bottom). The phase space selections are represented as black contours.}
\label{fig4}
\end{figure}

The signal K$^-$H $\rightarrow (\Sigma^{0}/\Lambda)\, \pi^{0}$ $if$ is characterized by an almost back-to-back production of the emerging $Y-\pi^0$ pairs. We use this feature to enhance the signal over background ratio. 
In Figs.~\ref{fig4} (top) and (bottom) the reconstructed MC $p_{\pi^0}$ versus $p_Y$ distributions are shown, for signal events in the $\Sigma^0 \pi^{0}$ and $\Lambda \pi^{0}$ channels, respectively. The simulations are used to optimize the two-dimensional cuts, which are represented in the Figs.~\ref{fig4} as black contours.

Additionally, thanks to the good resolution in the $\Lambda$ momentum ($\sigma_{p_\Lambda} \sim 1.9$~MeV/c for $\Lambda$s produced in the DC gas \cite{acta}), K$^-$H $\rightarrow \Lambda \pi^{0}$ ($ar$ and $if$) events can be effectively sampled as they are characterized by the sharp angular correlation $\cos\theta_{\Lambda \pi^0}<$~\mbox{-0.85} (where $\theta_{\Lambda \pi^0}$ is the angle between the two particles momentum vectors). Hence, this additional condition is set for the $\Lambda \pi^0$ sample.

\begin{figure}[h!]
\centering
\includegraphics[width=7.5cm,height=6.0cm]{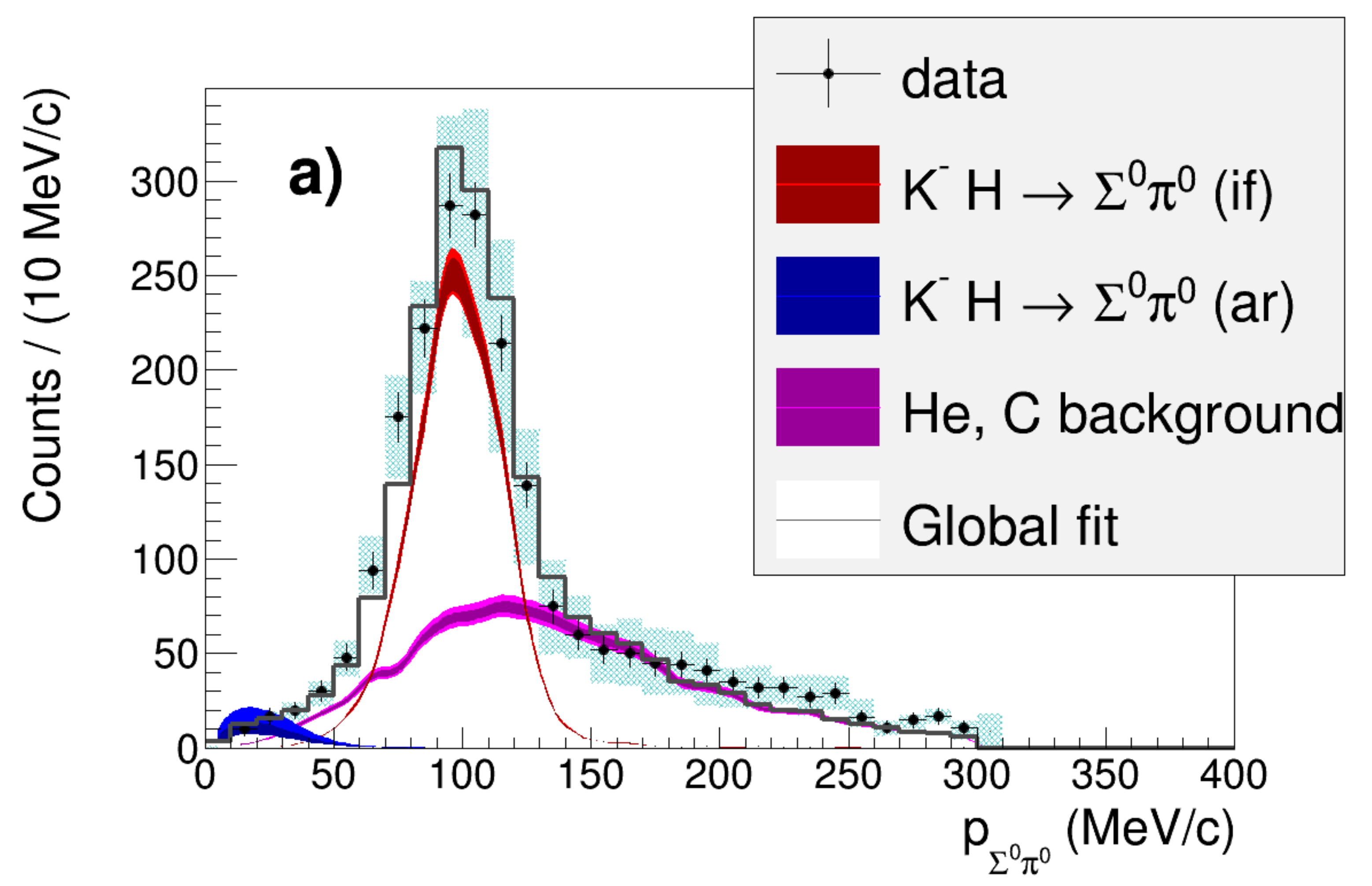}\\
\includegraphics[width=7.5cm,height=6.0cm]{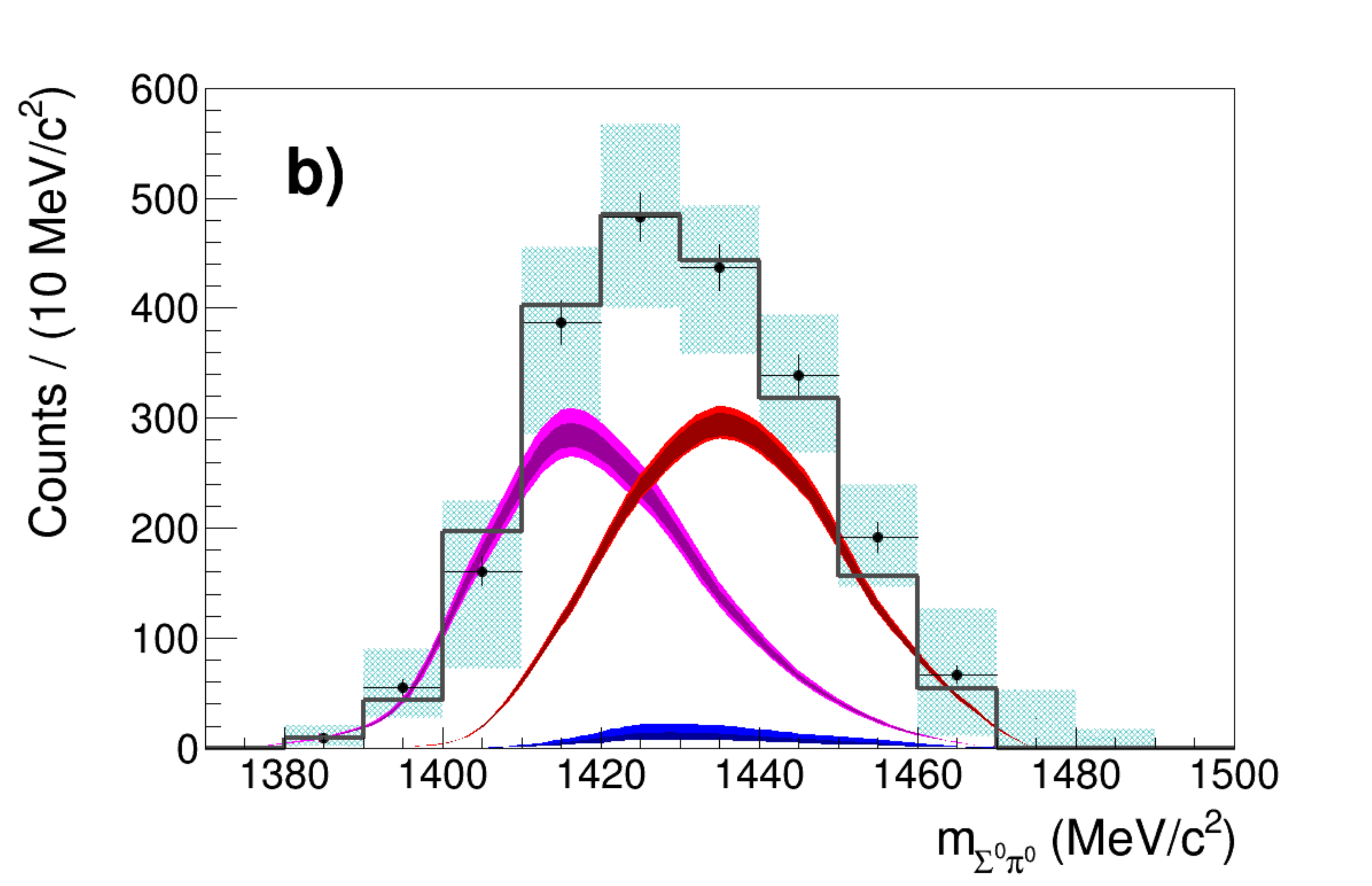}\\
\includegraphics[width=7.5cm,height=6.0cm]{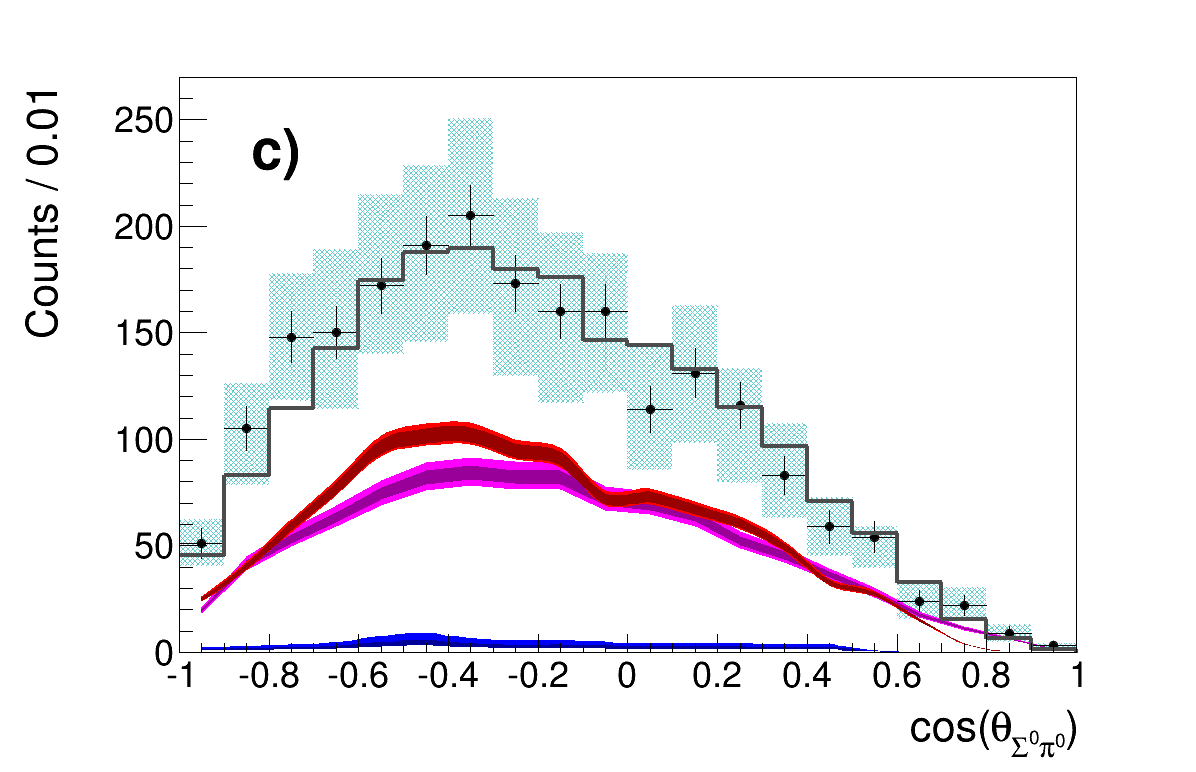}
\caption{
From top to bottom the figure shows the result of the simultaneous fit of p$_{\Sigma^{0}\pi^{0}}$, m$_{\Sigma^{0}\pi^{0}}$ and $cos\theta_{\Sigma^{0}\pi^{0}}$. The experimental data and the corresponding statistical errors are represented by black crosses, the systematic errors are light blue boxes. The contributions of the various physical processes are shown as colored histograms, according to the color code shown in the caption. The light and dark bands correspond to systematic and statistical errors, respectively. The gray distribution reproduces the global fit function. \label{finalfit1}}
\end{figure}

Panels a) in Figs. \ref{finalfit1} and \ref{finalfit2ll} exemplify the core of the final analysis step. Due to momentum conservation, the total hyperon-pion momentum ($p_{Y\pi^0}$) distributions of the K$^-$H $\rightarrow Y \pi^{0}$ $if$ samples reflect the original K$^-$ momentum spectrum, as demonstrated by the pre-eminent peaks centered at $p_{Y\pi^0} \sim 98$~MeV/c. This hallmark allows to efficiently disentangle the remaining background due to the competing K$^-$ absorption processes. 

At this stage of the events selection 6.5\% of the $\Sigma^{0} \pi^{0}$ events lie in the region $p_{\Sigma^0\pi^0}>300$~MeV/c, a phase space sector far away from the region of interest, mainly populated by FSI processes~\cite{lpi}. A final phase space selection $p_{\Sigma^0\pi^0}<300$~MeV/c is then applied. 

The selected $\Sigma^{0} \pi^{0}$ and $\Lambda \pi^{0}$ samples amount to 2130 and 7106 events, respectively. \newline

%\section{Cross sections determination}

%\subsection{Fit of the measured distributions}\label{fit}

The data are fitted by minimizing the following $\chi^2$ function:

\begin{equation}\label{defchi}
\chi^2 = \sum\limits_{q} \sum\limits_{n=1}^{N_{bins}^q} \frac{( N_n^q - \mathcal{F}^q (q_n) )^2}{\sigma_n^{q2}} \ ,
\end{equation}
in which $q$ ranges over the observables, $n$ over the number of bins of the $q$-th spectrum and $N_n^q$ is the measured content of the corresponding $n$-th bin. The fit function $\mathcal{F}$ contains the following physical processes (with the color code adopted in Figs.~\ref{finalfit1} and \ref{finalfit2ll}):

\begin{enumerate}

\item $K^- \, H \quad \rightarrow \quad (\Sigma^{0}/\Lambda) \,  \pi^{0}$ $if$ (red),

\item $K^- \, H \quad \rightarrow \quad (\Sigma^{0}/\Lambda) \, \pi^{0}$ $ar$ (blue),

\item $K^-  + \, {}^{4}He / {}^{12}C  \quad \rightarrow \quad \Sigma^{0}/\Lambda + \pi^{0} + {}^{3}H / {}^{11}B$ (magenta).

\end{enumerate}
In Ref. \cite{lpi} equivalent contributions are measured, within the uncertainties, for K$^-$ absorptions on $^4$He and $^{12}$C nuclei in the DC gas, as well as for the competing in-flight and at-rest nuclear interactions of the kaon. Accordingly, component 3 is prepared as an equal weight admixture of the mentioned reactions and the related uncertainty is considered as a systematic effect. For the $\Lambda \pi^0$ channel, besides the direct processes 1-3,  $\mathcal{F}$ also accounts for the same reactions initiated by $\Sigma^0 \pi^0$ production and followed by $\Sigma^0 \rightarrow \Lambda \gamma$ decay. $\mathcal{F}$ is given by the linear combination:
\begin{equation}
\mathcal{F}^q (q_n) = \sum\limits_{i=1}^{N_{par}} \alpha_i \cdot h_i^{ q} (q_n) \ ,
\end{equation}
with the index $i$ running over the number of free parameters $\alpha_i$ and $h_i^{ q}$ representing the distribution of the observable $q$, for the $i$-th physical process, normalized to the data entries. In Eq. \eqref{defchi} the errors are evaluated as $\sigma_n^{q  2} = \sqrt{\sum\limits_{i=1}^{N_{par}} (\alpha_i \cdot h_i^q(q_n))^2 + N_n^{q\ 2} }$ and the $\chi^2$ function minimization is performed by means of the SIMPLEX, MIGRAD and MINOS routines of ROOT \cite{MINUIT}.

For the $\Sigma^0 \pi^0$ sample a simultaneous fit is performed over $q = m_{\Sigma^0 \pi^0},\cos\theta_{\Sigma^0 \pi^0},p_{\Sigma^0 \pi^0}$, which indicate the $\Sigma^0 \pi^0$ invariant mass, angular correlation and total momentum, respectively. In the $\Lambda \pi^0$ channel the sharp selection of back-to-back events, renders the $\cos\theta_{\Lambda \pi^0}$ variable insensitive to the uncorrelated K$^-$ nuclear absorption reactions, hence the - kinematically equivalent - combination $q = m_{\Lambda \pi^0},p_{\Lambda}, p_{\pi^0}, p_{\Lambda \pi^0}$ is chosen. The fits results are presented in Fig.~\ref{finalfit1} (for $\Sigma^0  \pi^0$) and Fig.~\ref{finalfit2ll} (for $\Lambda  \pi^0$) and the obtained parameters values are summarized in Table~\ref{tablepar1}. 

\newpage

\begin{figure}[h!]
%\centering
\vspace{1.0cm}
\includegraphics[width=7.5cm,height=6cm]{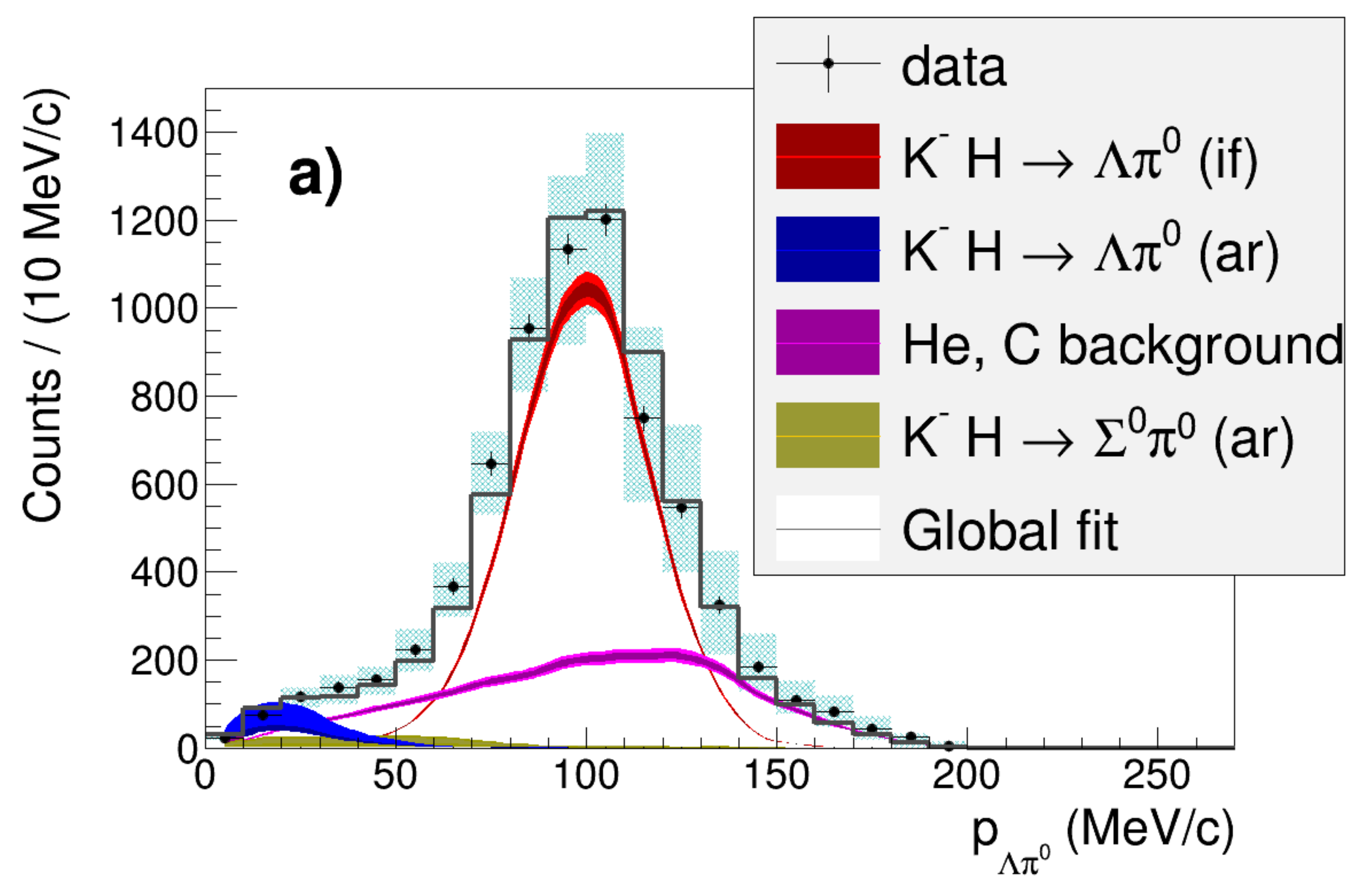}\\
\vspace{1.0cm}
\includegraphics[width=7.5cm,height=6cm]{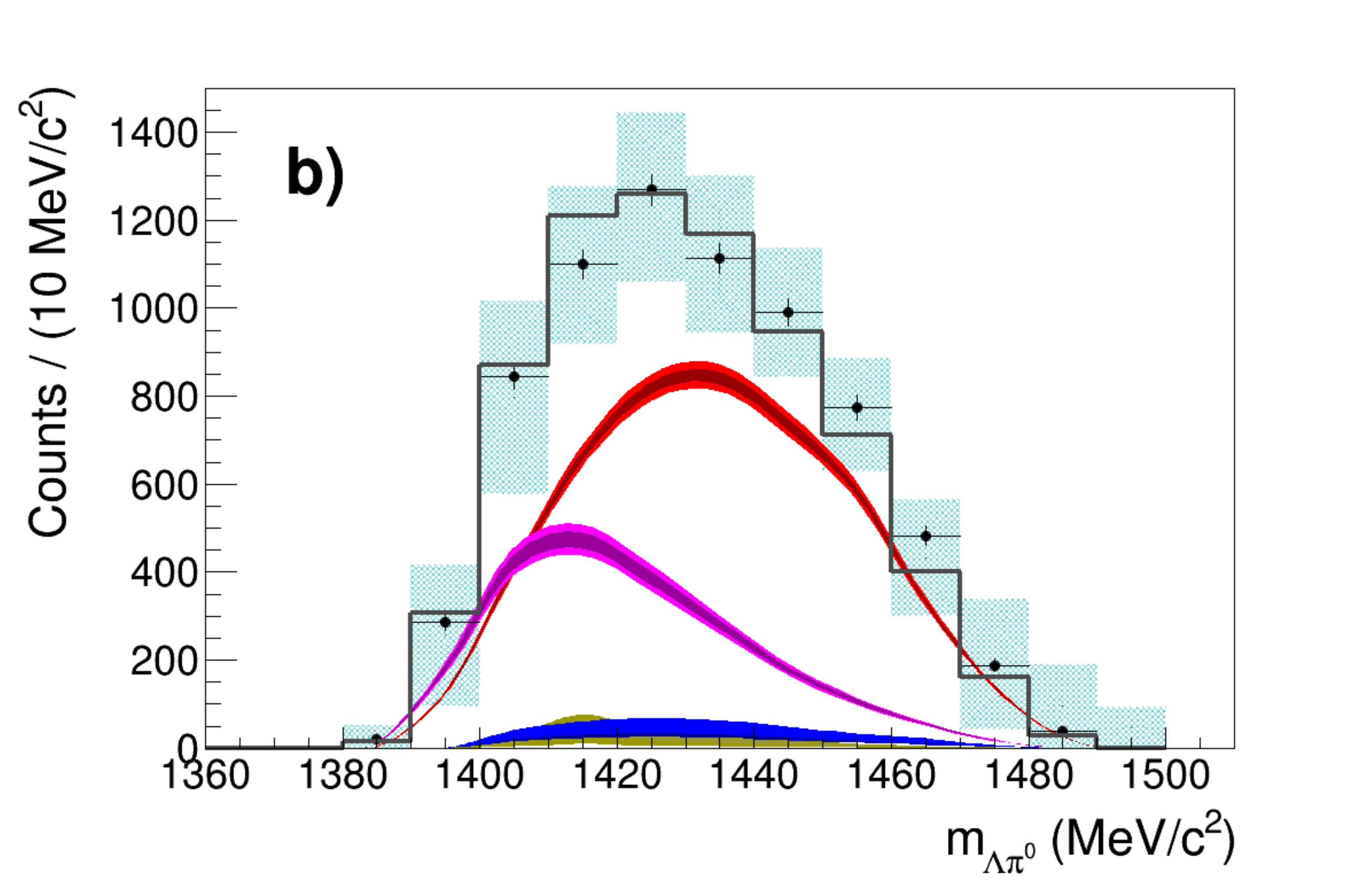}\\
\vspace{1.0cm}
\includegraphics[width=7.5cm,height=6cm]{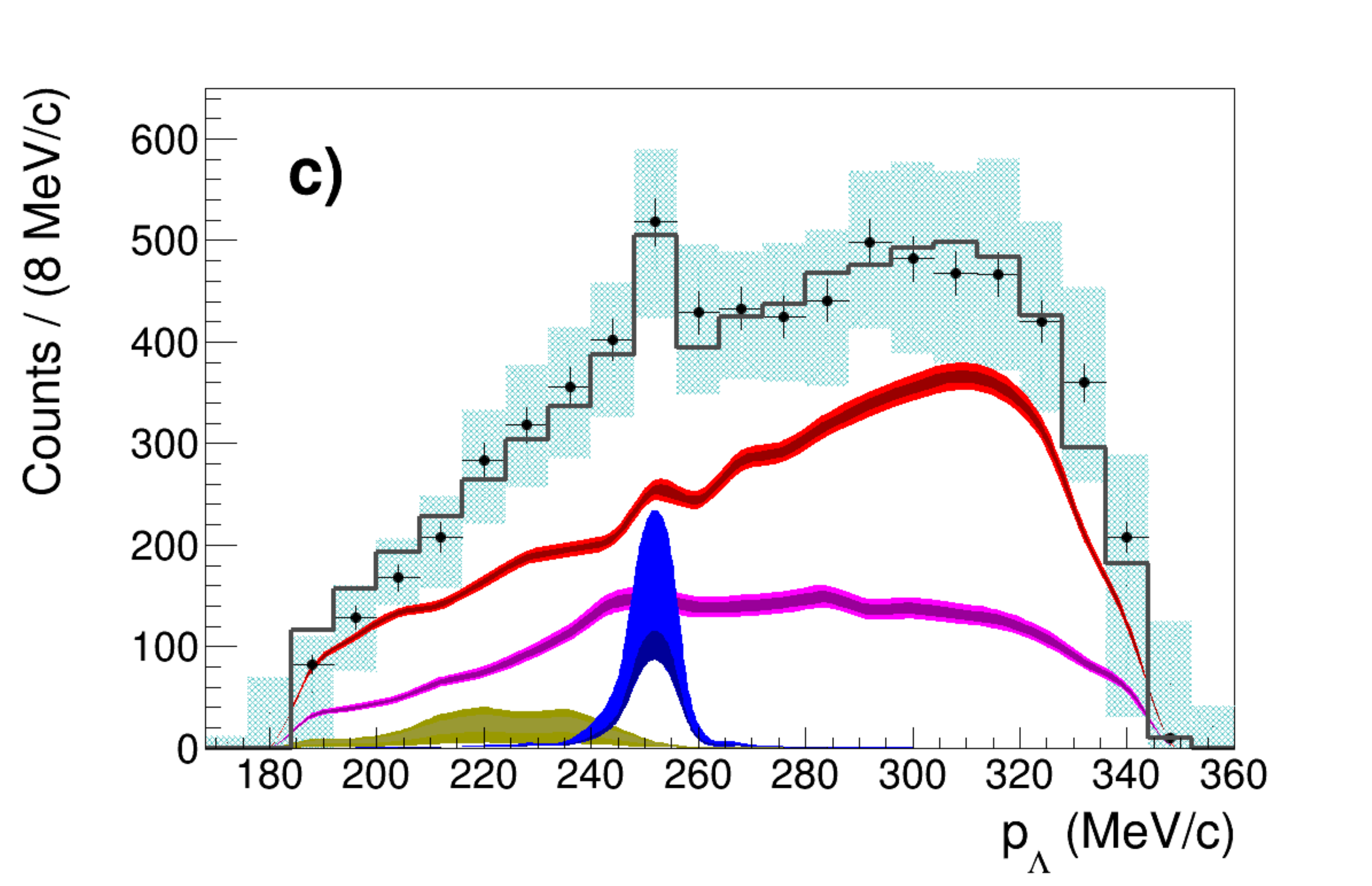}

\end{figure}

\begin{figure}[h!]
\includegraphics[width=7.5cm,height=6cm]{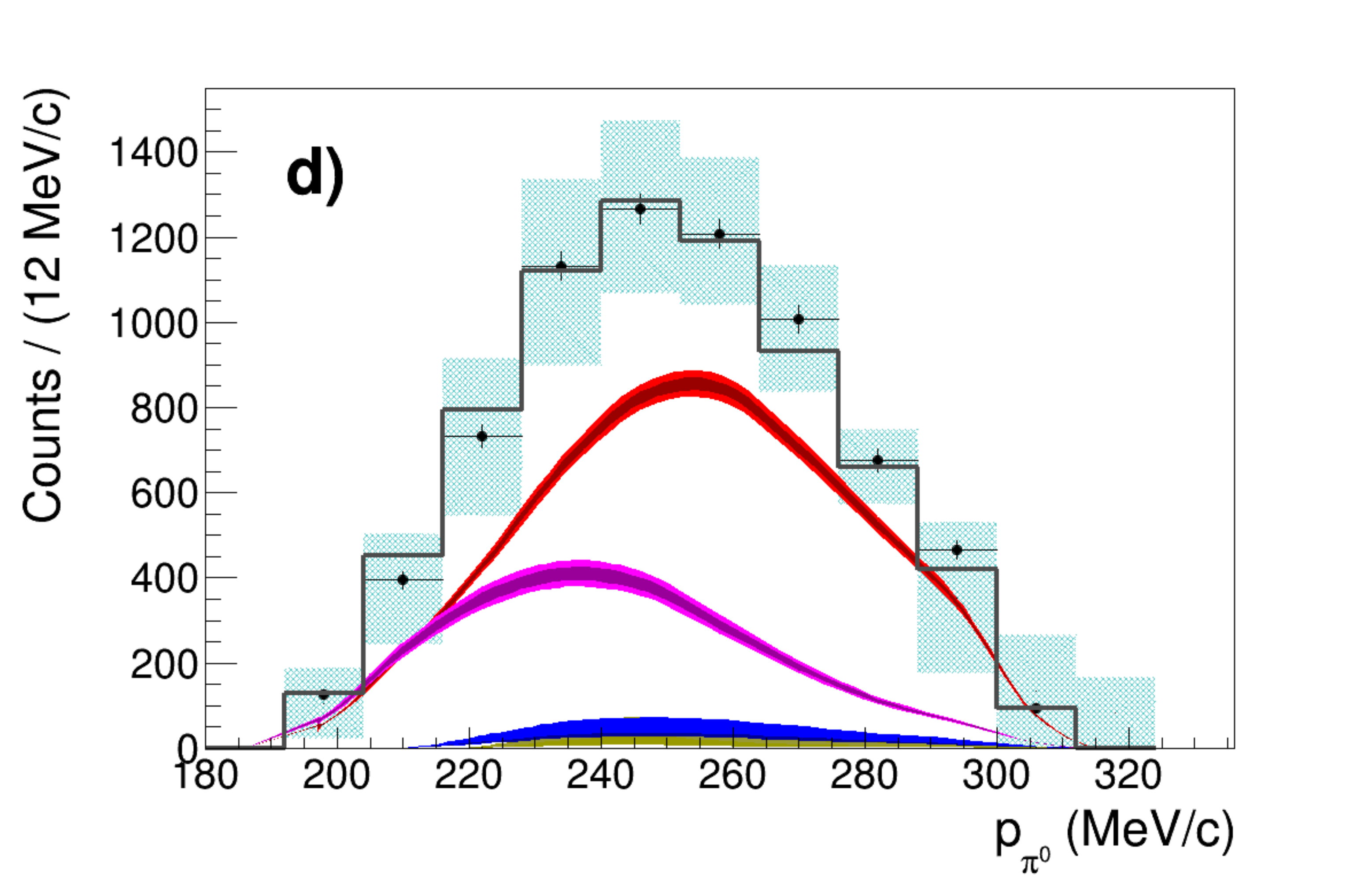}
\caption{From top to bottom the figure shows the result of the simultaneous fit of p$_{\Lambda\pi^{0}}$, m$_{\Lambda\pi^{0}}$, p$_{\Lambda}$ and p$_{\pi^{0}}$. The experimental data and the corresponding statistical errors are represented by black crosses, the systematic errors are light blue boxes. The contributions of the various physical processes are shown as colored histograms, according to the color code shown in the caption. The light and dark bands correspond to systematic and statistical errors, respectively. The gray distribution reproduces the global fit function. \label{finalfit2ll}}
\end{figure}

\newpage
The fit is not sensitive to the $\Lambda\pi^0$ kinematic distributions originated in $\Sigma^0$ decays, for the K$^- \, H$ interaction ($if$) and the nuclear K$^-$ absorptions; it yields negligibly small contributions, which are set to zero in the final fit.

The cross sections are calculated in accordance to:

\begin{equation}
\sigma=\frac{N^{if}_{K^-p \rightarrow Y\pi^{0}}}{N_{K^-}\cdot n \cdot L },
\end{equation}
where the total number of expected signal events $N^{if}_{K^-p \rightarrow Y\pi^{0}}$ is obtained re-scaling for the detection and reconstruction efficiencies and for the branching ratio of the $\Lambda \rightarrow p \pi^-$ decay.
Normalization to an absolute number of kaons is performed, taking advantage of the K$^+$ tagging~\cite{lproton,oton2016}. The number of projectiles $N_{K^-}$ corresponds to the number of tagged kaons corrected for the decay, by means of a MC simulation following the K$^-$ tracks (till the decay) in steps of 1 cm. $n$ is the number density of Hydrogen scattering centers. $L$ is the effective mean path length, obtained through a MC simulation, accounting for the K$^-$ impinging angle, i.e. the angle between the tangent vector to the particle’ trajectory at the DC entrance point, and the radial direction.

\begin{table}[!h]
\caption{The table summarizes the results obtained from the fits of the  $\Sigma^{0} \pi^{0}$ and $\Lambda \pi^{0}$ samples. The values of the reduced chi-squares and of the fit parameters are summarized.}
\centering
%{\setlength{\extrarowheight}{20pt}%
{\renewcommand{\arraystretch}{2}%
\begin{tabular}{cccc}
%\begin{tabular}{|l|l|}
\hline
\hline
$\bm{\Sigma^0 - \pi^0}$ \textbf{CHANNEL} & 
$\frac{\chi^2}{dof} = \frac{92}{54}$& &\\
\hline 
\textbf{process} & \textbf{fit par. value} & $\mathbf{\sigma_{stat.}}$ \\
%\vspace{0.05cm}
\hline
%\vspace{0.05cm}
$K^-  H \rightarrow  \Sigma^{0}  \pi^{0} \, (if)$ & 0.511 &	$\pm$ 0.018	\\
$K^-  H \rightarrow \Sigma^{0}  \pi^{0} \, (ar)$ & 0.017	& $\pm$ 0.005 \\
$K^- + ^{4}He/^{12}C \rightarrow \Sigma^{0} \pi^{0}$ & 	&  \\

+ residual $(ar/if)$ & 0.463	& $\pm$ 0.018 \\

\hline
\hline
$\bm{\Lambda - \pi^0}$ \textbf{CHANNEL} & 
$\frac{\chi^2}{dof} = \frac{165}{57}$& &\\
\hline 
\textbf{process} & \textbf{fit par. value} & $\mathbf{\sigma_{stat.}}$ \\
%\vspace{0.05cm}
\hline

$K^-  H \rightarrow  \Lambda  \pi^{0} \, (if)$ & 0.659 &	$\pm$ 0.011	\\
$K^-  H \rightarrow \Lambda  \pi^{0} \, (ar)$ & 0.021	& $\pm$ 0.003 \\
$K^- + ^{4}He/^{12}C \rightarrow \Lambda \pi^{0}$ & 	&  \\

+ residual $(ar/if)$ & 0.298	& $\pm$ 0.012 \\

$K^-  H \rightarrow  \Sigma^0  \pi^{0}$  &  &		\\

$\rightarrow  \Lambda \gamma  \pi^{0} \, (ar)$ & 0.018 &	$\pm$ 0.006	\\

\end{tabular}}
\label{tablepar1}
\end{table}

The obtained cross sections 
\begin{itemize}
    \item $\sigma_{K^- p \rightarrow \Sigma^0 \pi^0} = 42.8 \pm 1.5 (stat.) ^{+2.4}_{-2.0}(syst.) \ \mathrm{mb}$
    \item $\sigma_{K^- p \rightarrow \Lambda \pi^0} = 31.0 \pm 0.5 (stat.) ^{+1.2}_{-1.2}(syst.) \ \mathrm{mb}\,$,
\end{itemize}
correspond to a mean kaon momentum $p_K = (98 \pm 10)$~ MeV/c, calculated on the basis of the true MC information of the kaons momentum distribution inside the DC volume.

The systematic errors are determined repeating several times the same fit procedure, by varying independently all the analysis cuts which were optimized for the $\Sigma^0 \pi^0$ and $\Lambda \pi^0$ samples selection. The systematic error on the $i$-th parameter of the fit, due to a variation of the $j$-th cut, is defined as:

\begin{equation}
\sigma_{syst,i}^{j}=\alpha_{i}^{j}-\alpha_{i}
\end{equation}
Total, positive and negative systematic errors are obtained by summing in quadrature the positive and negative systematic fluctuations.

With the exception of those quantities for which the statistical error is known (e.g. $m_{\gamma_1 \gamma_2}$, $m_{\Lambda \gamma3}$ and $\cos\theta_{{\Lambda \pi^0}}$), in which case the systematics are evaluated by applying $\pm 1\sigma$ fluctuations to the corresponding cuts, 
%and of the background sources whose contribution is known by simulations (e.g. the background introduced by the $\rho_\Lambda$ cut), 
for the other selections we chose to change the cuts of the amount necessary to increase (or decrease) the selected number of events of 15\%, with respect to the standard. This is the case of the constraints on $\rho_\Lambda$, $\chi_t^2$, $\chi^2_{m_{\gamma_1\gamma_2}}$ and $\chi^2_{m_{\Lambda\gamma_3}}$, of the phase space selections in the $p_{\pi^{0}}-p_{\Sigma^{0}}$ and in the $p_{\pi^{0}}-p_{\Lambda}$ planes.
The systematic uncertainty introduced by setting equal contributions of K$^-$ absorptions on Helium and Carbon, both for the $ar$ and $if$ processes, is set by performing 15\% variations of the relative contribution of each process.

The $p_{\Sigma^{0}\pi^{0}}$ constraint was optimized based on a scan in the range ($280\div350$)~MeV/c, in steps of 10~MeV/c (compatible with the resolution $\sigma_{p_{\Sigma^0\pi^0}} \sim 15$~MeV/c) yielding the minimum reduced $\chi^2$ for $p_{\Sigma^{0}\pi^{0}}=300$~MeV/c. The contribution to the systematic errors is obtained by the condition $p_{\Sigma^{0}\pi^{0}}<310$~MeV/c.

The systematics introduced by the decay correction in the $N_{K^-}$ calculation and by the evaluation of $L$ are estimated by doubling the 1 cm step length, and diminishing of 15\% the number of simulated kaons. \newline

%It is to be stressed that this work gives the first direct measurement of the low-energy $K^- p \rightarrow \Sigma^0 \pi^0$ cross section. The other experimental points, shown in Fig. \ref{comparison1}, are indeed obtained in Refs. \cite{s0pi01,s0pi02} indirectly, from the measurement of the $K^- p \rightarrow \Lambda \pi^0$ process, assuming the isospin invariance.

%In case of the cut on the $m_{\Lambda \gamma 3}$ - $m_{\gamma 1\gamma 2}$ scatterplot, the semi-axes of the ellipse were increased and decreased by $\pm 1\sigma_{\Lambda \gamma_3}$ (the semi-minor axis) and $\pm 1\sigma_{\gamma_1 \gamma_2}$ (the semi-major axis). 

%\section{Conclusions}

We reported the analysis, by the AMADEUS collaboration, of the 1.74 fb$^{-1}$ integrated luminosity collected by the KLOE collaboration in the period 2004/2005. Taking advantage of the low momentum K$^-$ beam delivered by the DA$\Phi$NE collider, two independent and high statistics samples of K$^-$ absorptions in-flight on Hydrogen were disentangled, with final states $\Sigma^0\pi^0$ and $\Lambda \pi^0$, which allowed to perform high precision measurements of the inelastic K$^{-}$p $\rightarrow (\Sigma^0/\Lambda)\,\pi^{0}$ cross sections, for a K$^{-}$ momentum $p_K = (98 \pm 10)$~MeV/c:

\begin{itemize}
    \item $\sigma_{K^- p \rightarrow \Sigma^0 \pi^0} = 42.8 \pm 1.5 (stat.) ^{+2.4}_{-2.0}(syst.) \ \mathrm{mb}$
    \item $\sigma_{K^- p \rightarrow \Lambda \pi^0} = 31.0 \pm 0.5 (stat.) ^{+1.2}_{-1.2}(syst.) \ \mathrm{mb}\,$.
\end{itemize}
With respect to previous experiments (\cite{s0pi01,martin,ferrari} and references therein), which extrapolated $\sigma_{K^- p \rightarrow \Sigma^0 \pi^0}$ from the measurement of the K$^{-}$p $\rightarrow \Lambda\pi^{0}$ cross section, assuming isospin symmetry, this is the first direct simultaneous measurement of the cross sections in the isospin $I=1$, and almost pure $I=0$ channels. 
Alongside a comparable range for the kaon momentum ($\pm$10~MeV/c with respect to $\pm$12.5~MeV/c in Refs. \cite{s0pi01,martin,ferrari} and references therein)
the precision achieved in the cross section determination greatly overcomes all the other low momentum measurements, 
providing a key input to the theory for the extrapolation towards and below the $\overline{K}N$ threshold region.

%The measured cross sections are characterised by much smaller errors compared to Refs.~\cite{s0pi01,s0pi02} and are barely consistent with the theoretical prediction Ref.~\cite{IHW}. It is although important to stress that the fit of the coupled-channels dynamical calculation based on the next-to-leading order chiral SU(3) meson–baryon effective Lagrangian presented in Ref. \cite{IHW} did not account for the cross sections ranges on $p_{K^{-}}$.

%This represents a key information for the non-perturbative chiral SU(3) coupled channels models, since it allows self-consistent extrapolation to the below threshold region, where the $\overline{K}N \rightarrow Y\pi$ coupled dynamics is dominated by these two channels.

\section*{Acknowledgements}
We acknowledge the KLOE/KLOE-2 Collaboration for their support and for having provided us with the data and the tools to perform the analysis presented in this paper. Part of this work was supported by the EU STRONG-2020 project (grant agreement No 824093); 
Minstero degli Affari Esteri e della Cooperazione
Internazionale, EXOTICA project, PO21MO03;
the Polish National Science Center through grant No. UMO-2016/21/D/ST2/01155.
The project is co-financed by the Polish National Agency for Academic Exchange, grant no PPN/BIT/2021/1/00037.
The authors acknowledge support from the SciMat and qLife Priority Research Areas budget under the program Excellence Initiative-Research University at the Jagiellonian University.

%\bibliography{../LP/LP}
%\bibliographystyle{elsarticle-num}

\bibliographystyle{unsrt}
\end{document}